\documentclass[a4paper,10pt,twoside]{cpc-hepnp}

\usepackage{multicol}
\usepackage{graphicx}
\usepackage{booktabs}
\usepackage{amssymb,bm,mathrsfs,bbm,amscd}
\usepackage[tbtags]{amsmath}
\usepackage{lastpage}
\usepackage{enumerate}

\begin{document}

\fancyhead[c]{\small Chinese Physics C~~~Vol. xx, No. x (2014) xxxxxx}
\fancyfoot[C]{\small 010201-\thepage}

\footnotetext[0]{}

\title{Measurement of the dead layer thickness in a p-type point contact germanium detector}

\author{%
      H. Jiang$^{1,2}$\
\quad Q. Yue$^{1,2;1)}$\email{yueq@mail.tsinghua.edu.cn}%
\quad Y.L. Li$^{1,2;2)}$\email{yulanli@mail.tsinghua.edu.cn}%
\quad K.J. Kang$^{1,2}$
\quad Y.J. Li$^{1,2}$
\quad J. LI$^{1,2}$
\quad S.T. Lin$^{3}$
\quad S.K. Liu$^{1,2}$
\\H. Ma$^{1,2}$
\quad J.L. Ma$^{1,2}$
\quad J. Su$^{1,2}$
\quad H.T. Wong$^{4}$
\quad L.T. Yang$^{1,2}$
\quad W. Zhao$^{1,2}$
\quad Z. Zeng$^{1,2}$
}
\maketitle

\address{%
$^1$ Dept. of Engineering Physics, Tsinghua University, Beijing 100084, China\\
$^2$ Key Laboratory of Particle $\&$ Radiation Imaging (Tsinghua University), Ministry of Education, Beijing 100084, China\\
$^3$ College of Physical Science and Technology, Sichuan University, Chengdu 610064, China\\
$^4$ Institute of Physics, Academia Sinica, Taipei 11529, China\\
}

\begin{abstract}
A 994g mass p-type PCGe detector was deployed by the first phase of the China Dark matter EXperiment aiming at the direct searches of light weakly interacting massive particles.  Measuring the thickness of the dead layer of a p-type germanium detector is an issue of major importance since it determines the fiducial mass of the detector. This work reports a method using an uncollimated $^{133}$Ba source to determine the dead layer thickness. The experimental design, data analysis and Monte Carlo simulation processes, as well as the statistical and systematic errors are described. An agreement between the experimental data and simulation results was achieved to derive the thickness of the dead layer of 1.02 mm.
\end{abstract}



\footnotetext[0]{\hspace*{-3mm}\raisebox{0.3ex}{$\scriptstyle\copyright$}2013
Chinese Physical Society and the Institute of High Energy Physics
of the Chinese Academy of Sciences and the Institute
of Modern Physics of the Chinese Academy of Sciences and IOP Publishing Ltd}%

\begin{multicols}{2}

\section{Introduction}

The China Dark matter EXperiment (CDEX) aims at the direct searches of light Weakly Interacting Massive Particles (WIMPs) employing point-contact germanium detector (PCGe) at the China Jinping Underground Laboratory (CJPL), which has about 2400 m of rock overburden. As the first step, CDEX has reported the results from the CDEX phase I experiment (CDEX-1) by using a p-type PCGe (PPCGe) detector (CDEX-1A) of mass 994 g~\cite{lab1}. PPCGe detector has very excellent properties for the dark matter search experiment. A small area of the point contact electrode, which can significantly reduce the capacitance, results in the low electronic noise and energy threshold~\cite{lab2}. The CDEX-1A detector can reach a threshold of $\sim$400 eVee (¡°ee¡± denotes electron-equivalent energy)~\cite{lab3}. Besides that, the localized weighting potential resulting in distinct current pulses from individual interaction charge clouds provides the ability to distinguish between Single Site Events (SSE) and Multiple Site Events (MSE) which can be used to reduce the signals from the background~\cite{lab4}.

However, PPCGe detector has a dead layer at the surface caused by lithium diffusion. Events generating in the dead layer will lead to a slow rise time pulse and an incomplete charge collection because of the very weak electric field in this region~\cite{lab5,lab6,lab7}. Burns et al. found that the dead layer could be considered as two layers: an inactive layer where no charge could be collected by the electrodes, and a transition layer where the charge collection efficiency increased from zero to one~\cite{lab8}. As a result, the events in the dead layer cannot provide us the primary energies that deposited in the detector and should be discriminated from the bulk events with efficiency correction. At the same time, the fiducial volume and mass should also be calculated based on the thickness of the dead layer.

The characteristics of the dead layer have been investigated in some literatures.  In 1998, Clouvas used a formula to describe the charge collection in the dead layer. There was a good agreement between the simulation spectrum and experimental spectrum based on this formula. But they got a dead layer thickness of 2.5 mm which had a huge disagreement with the manufactory's result of 0.5 mm~\cite{lab9}. Studies of N. Q. Huy et al. showed that the dead layer thickness increased over time because the lithium diffusion was happening all the time~\cite{lab10}. Thus, it is not reliable to use the dead layer thickness given by manufactory. In order to calculate the fiducial volume and mass, the thickness of the dead layer should be measured.

This article reports a measurement of the dead layer thickness of the PPCGe detector from CDEX-1A experiment with an uncollimated $^{133}$Ba source. In the following sections, the experimental design, data analysis and Mote Carlo simulation processes, as well as the statistical and systematic errors are discussed.

\section{Detector geometry}

The CDEX-1A detector was fabricated by Canberra Company. The germanium crystal has a diameter of 62.2 $\pm$ 0.1 mm and a height of 62.3 $\pm$ 0.1 mm and is encapsulated in a Oxygen Free High Conductivity (OFHC) copper cryostat. The copper thickness at the endcap side opposite of the point contact is 1.5 $\pm$ 0.1 mm and at the side is 2.0 $\pm$ 0.1 mm. The distance between the endcap and the germanium crystal is 4.5 $\pm$ 0.1 mm and no other material but vacuum between them. At the side, there are some supports, foils and screws made of OFHC copper, lead and brass. The point contact is at the bottom of the crystal and connected to a brass pin to take out the signals.

The PPCGe detector can be distinguished into two parts: a dead layer where no charge or part of the charges is collected by electrodes, a bulk where all the charges deposited in it can be collected by electrodes. Signals generating in the dead layer have a much slower rise time and an incomplete charge collection as shown in Fig.~\ref{fig1}.

\begin{center}
  \includegraphics[width=8.0cm,height=6.0cm]{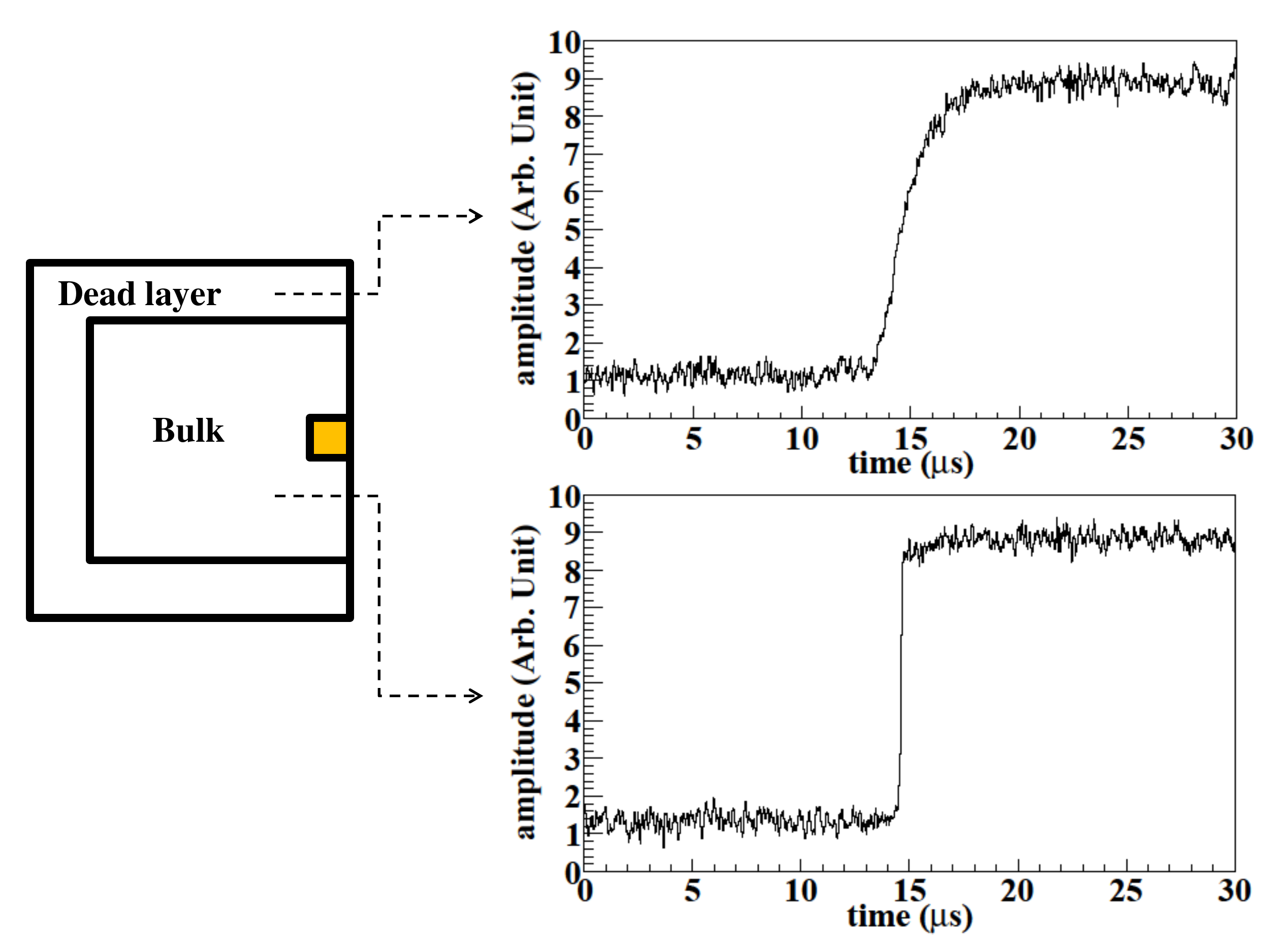}
  \figcaption{\label{fig1}Typical signals at $\sim$10 keV from dead layer (top) and bulk (bottom) . Signals from dead layer have a much slower rise time and an incomplete charge collection, e.g. the actual deposit energy is larger than 10 keV. The size of the point and the thick of the dead layer are not shown in scale. }
\end{center}

\section{Experimental process}

A $^{133}$Ba source is an excellent source to measure the dead layer, which has several photoelectron peaks from low to high energy as shown in Table~\ref{tab1}. Lower energy photons are more likely to interact in the surface comparing to the higher energy photons and due to the partial charge collection, the surface events do not contribute to the photoelectron peaks at all. It means that different dead layer thicknesses will cause different intensity ratios of photoelectron peaks for the gamma rays with different energies.

\begin{center}
\tabcaption{ \label{tab1}   Gamma ray intensities of a $^{133}$Ba source~\cite{lab11}}
\footnotesize
\begin{tabular*}{50mm}{c@{\extracolsep{\fill}}cc}
\toprule Energy/keV  && Intensity/$\%$\\
\hline
53.16 && 2.14 \\
79.61 && 2.65 \\
81.00 && 32.95 \\
160.61 && 0.64 \\
223.24 && 0.45 \\
276.40 && 7.16 \\
302.85 && 18.34 \\
356.01 && 62.05 \\
383.85 && 8.94 \\
\bottomrule
\end{tabular*}
\end{center}

The steps to measure the dead layer thickness are following: Firstly, an uncollimated $^{133}$Ba source was used to get the ratios of different photoelectron peaks in experiment. Secondly, as all the parameters except the dead layer thickness were known, a series of dead layer thicknesses were assumed to get the simulation results of the ratios, respectively. At last, the dead layer thickness was derived by comparing the experimental data and simulation data. This method has also been used by GERDA and MAJORANA Collaborations to get the dead layer thickness of their germanium detectors~\cite{lab12}.

\begin{center}
  \includegraphics[width=7.2cm,height=5.4cm]{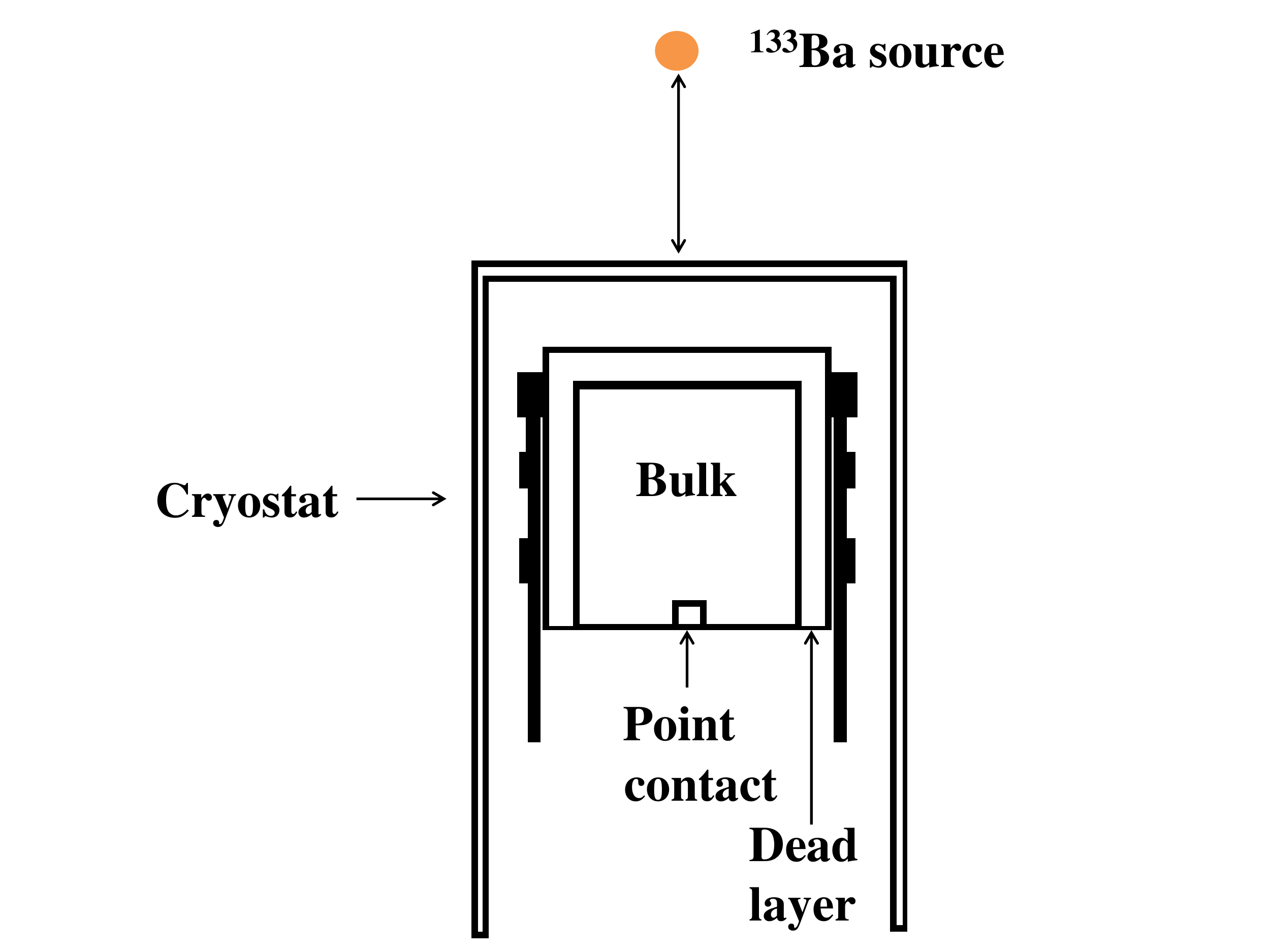}
  \figcaption{\label{fig2}The layout of the CDEX-1A PPCGe detector and source position. The point and dead layer are not shown in scale. }
\end{center}

In the experiment, an uncollimated $^{133}$Ba source was put right above the detector endcap, as shown in Fig.~\ref{fig2}. A height gauge was used to fix the $^{133}$Ba source and measure the distance between the source and the endcap. Signals from the point contact electrode were fed into a pulsed reset preamplifier. Then, a shaping amplifier at 6us shaping time and a 14 bit 100 MHz flash analog-to-digital convertor (FADC) were used to shape, amplify and digitize the signals.

\section{Results and discussion}

\subsection{Experimental data analysis}

As the experiment is based on the contrast of experimental results and simulation results, it is very necessary to make sure that the experimental conditions are quite the same with the simulation conditions. In order to achieve this goal, several events selections and rejections should be done in the experimental data before we can get the experimental spectrum of a $^{133}$Ba source.

\subsubsection{Events selection from preamplifier reset period}

A pulsed reset preamplifier is used by CDEX-1A PPCGe detector to achieve ultra-low noise level. The charge and discharge procedures are shown in Fig.~\ref{fig3}(a). The baseline level of the preamplifier decreases with the time due to the continuous leakage current of the detector itself and the induced current (signal) from the incident particles. This is the so called charge procedure. When the baseline level reaches to the reset point, the preamplifier would be reset immediately to make it work again. This is the so called discharge procedure.

\begin{center}
  \includegraphics[width=8.45cm,height=7.27cm]{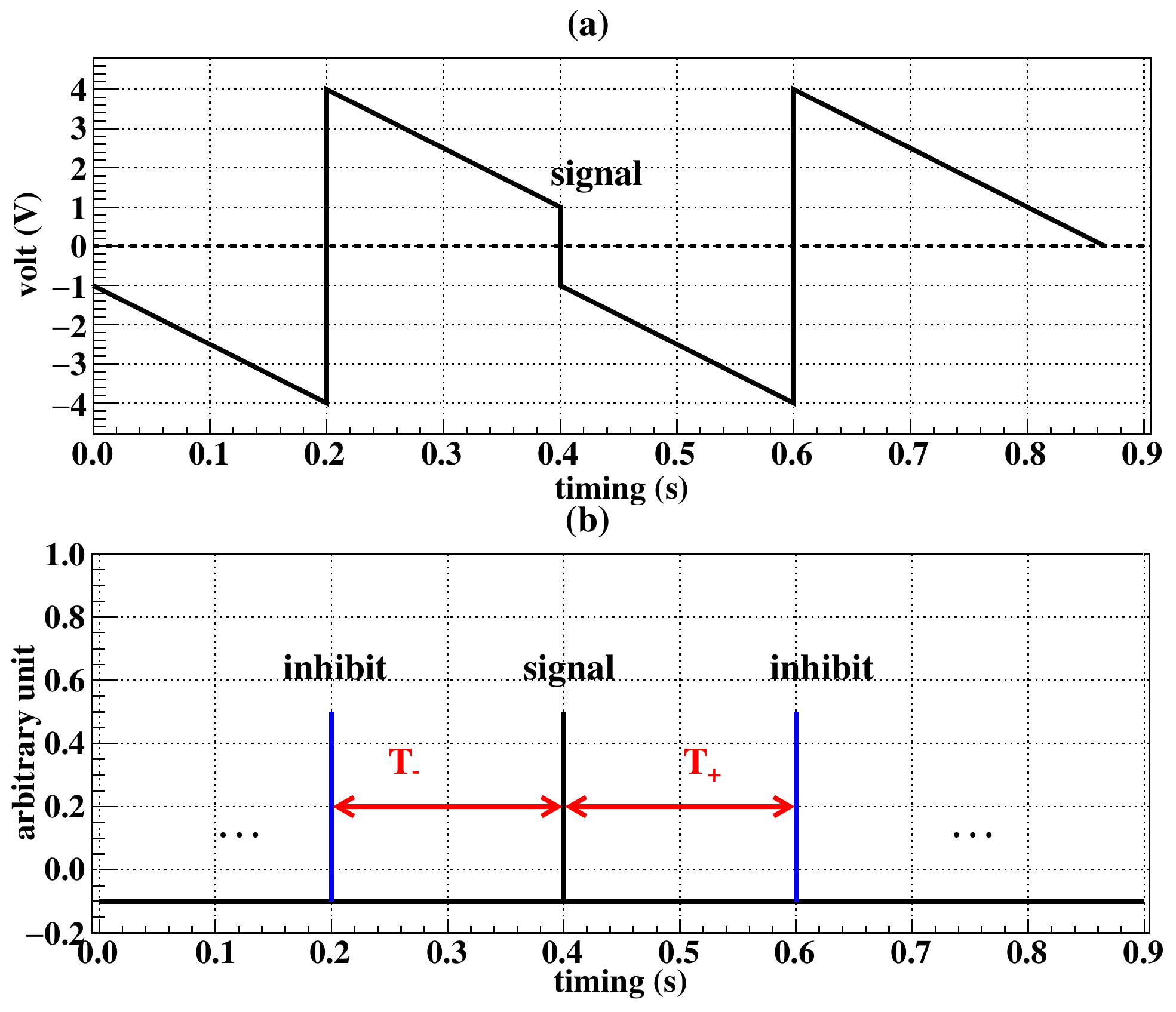}
  \figcaption{\label{fig3}(a) Baseline level of the preamplifier. (b) Each time the preamplifier is reset, a reset inhibit signal is obtained and T$_{-}$ means the time between the event to its nearest prior reset inhibit signal.}
\end{center}

Each time the preamplifier is reset, a reset inhibit signal is obtained as shown in Fig.~\ref{fig3}(b) and the typical average reset period is about 0.4 s with a $^{133}$Ba source irradiation for CDEX-1A PPCGe detector. T$_{-}$ means the time difference between the event and its nearest prior reset inhibit signal.

If the electronic level difference between the baseline level to the reset point is less than the voltage drop produced by a large induced current from an incident particle, the output signal just represents part of the total energy deposited in the detector. This effect decreases the detection efficiency for the events of the single-energy incident gamma-rays. And the higher energy particles are more easier to reset the preamplifier. That means the detection efficiency is energy dependent which will change the ratios of different photoelectron peaks.

The events selection from preamplifier reset period was derived from the parameter  T$_{-}$ distribution, aiming at choosing a region where the detection efficiency was energy independent to make sure that the ratios of different photoelectron peaks were not affected. Fig.~\ref{fig4} shows the relationship between the event counts of a $^{133}$Ba source and T$_{-}$ at the source position of 73 mm. It is shown that the event counts were dropped rapidly at T$_{-}$ $>$ 0.22 s which meant some gamma rays were lost as the preamplifier was reset by them. In order to avoid the efficiency problem, we set T$_{-}$ cut at 0.2 s to reject the events with T$_{-}$ $>$ 0.2 s.

\begin{center}
  \includegraphics[width=8.45cm,height=5.65cm]{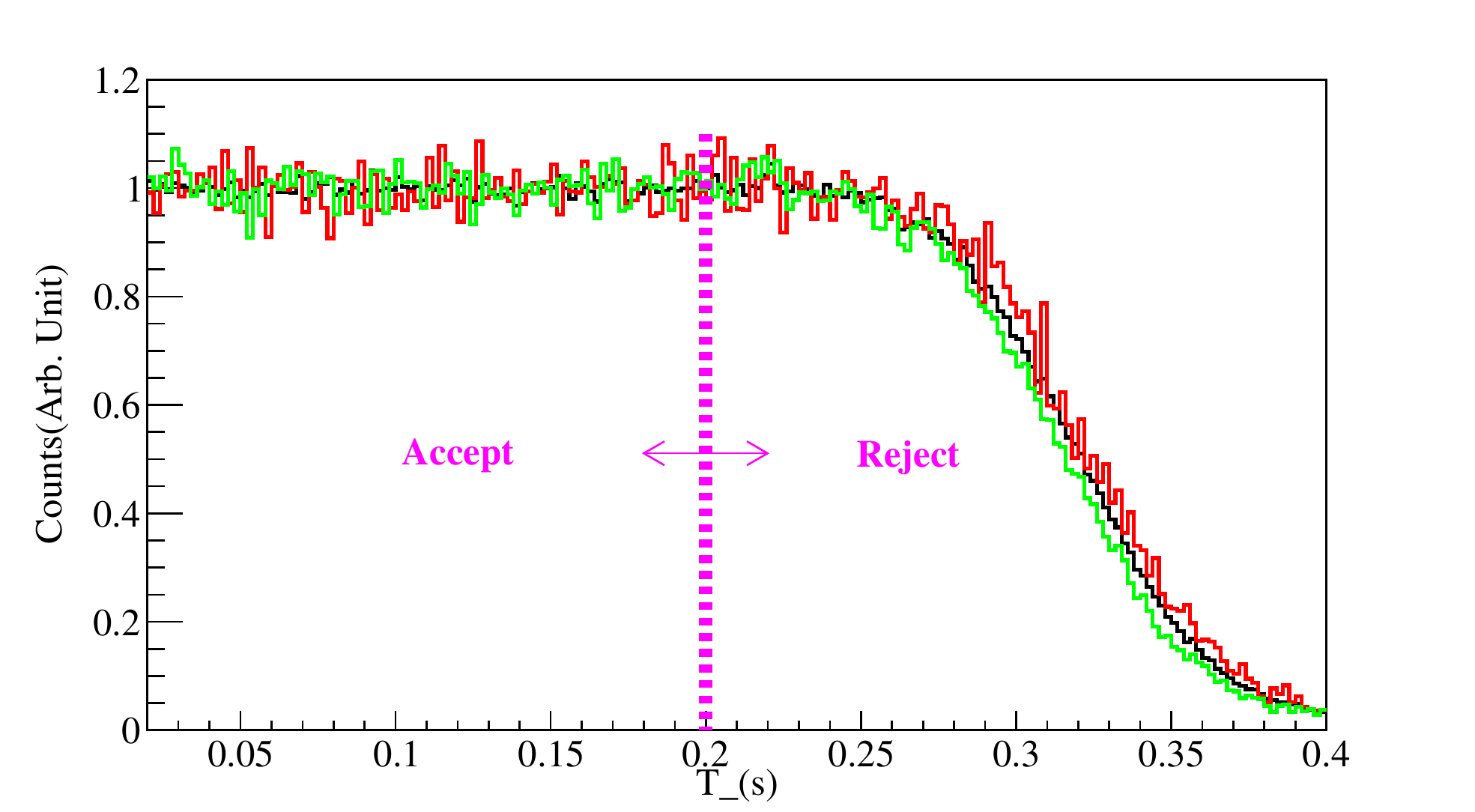}
  \figcaption{\label{fig4}Relationship between event counts and T$_{-}$ at source position 73 mm. The black line is the event counts of the total spectrum. The red line is the event counts of the 81 keV photoelectron peak. The green line is the event counts of the 356 keV photoelectron peak. The counts for photoelectron peaks of 81 keV, 356 keV and the total spectrum have been normalized to the same scale in this figure.}
\end{center}

\subsubsection{Accidental coincidence events rejection}

There are some accidental coincidence events in the time window of our DAQ system. Since the amplitude was used to do the energy calibration, some lower energy events could be masked by higher energy events which would also change the ratios of different energy peaks. The rejection criterion of accidental coincidence events is based on the correlation between the maximal amplitude(Amp) and the integral of the signal pulse(Q).

Fig.~\ref{fig5} shows the relationship between Q and Amp. The main distribution band are normal events because the Q-Amp distribution are linear correlation, while the accidental coincidence events display different behaviors. The signal selection region is defined at 3$\sigma$ of the linear events distribution band.

\begin{center}
  \includegraphics[width=8cm,height=4.4cm]{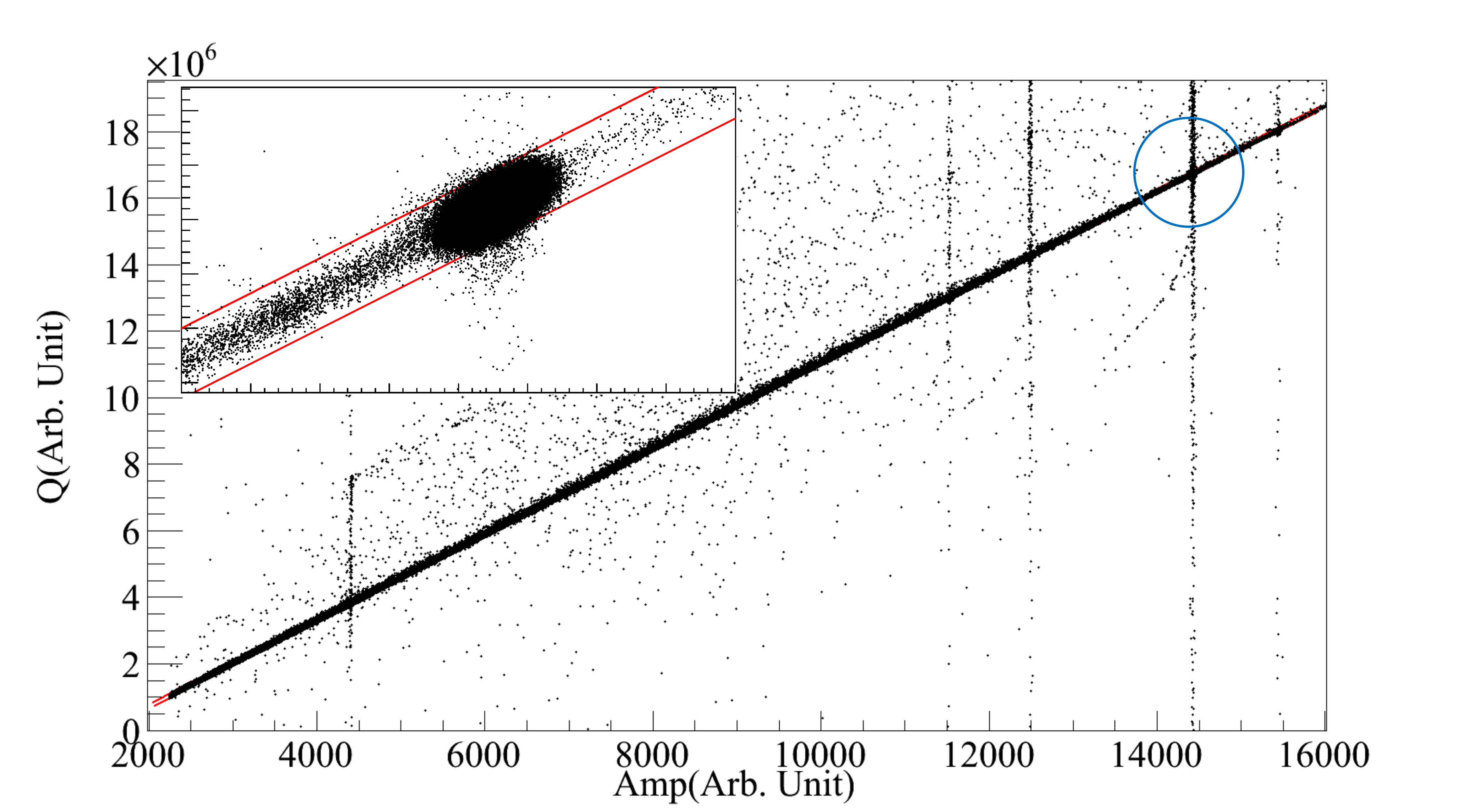}
  \figcaption{\label{fig5}All the events outside the two red lines are rejected. The inset figure shows the 356 keV photoelectron peak and the 3$\sigma$ cut lines.}
\end{center}

The relationship between the real rate and measured rate of one certain energy photoelectron peak is shown as the formula below:

 \begin{eqnarray}
\label{test}
R_{m} = R_{r}-2{\tau}{R_{r}}{R_{tot}}=R_{r}(1-2{\tau}R_{tot})
\end{eqnarray}

Where $R_{m}$ refers to the measured rate of the certain energy photoelectron peak after the accidental coincidence events rejection,  $R_{r}$ refers to the real rate of the certain energy photoelectron peak detected by the detector, $\tau$ refers to the time window of the accidental coincidence, $R_{tot}$ refers to the real event rate of the whole spectrum detected by the detector.

As the parameter $1-2{\tau}R_{tot}$ in formula (1) is a constant term which is energy independent, it would not change the value of ratios of different photoelectron peaks. That means the ratios of measured rates after accidental coincidence events rejection are equal to the ratios of real rates. We can use the measured rates to derive the ratios of different photoelectron peaks without any correction.

\subsubsection{Experimental results}

After doing all the selections and rejections, an experimental spectrum of a $^{133}$Ba source is shown in Fig.~\ref{fig6}. The distance between the source and endcap is 73 mm. In this spectrum, there were five explicit photoelectron peaks: 81keV, 276 keV, 303 keV, 356 keV and 384 keV which were used to obtain the dead layer thickness.

\begin{center}
  \includegraphics[width=8cm,height=4.4cm]{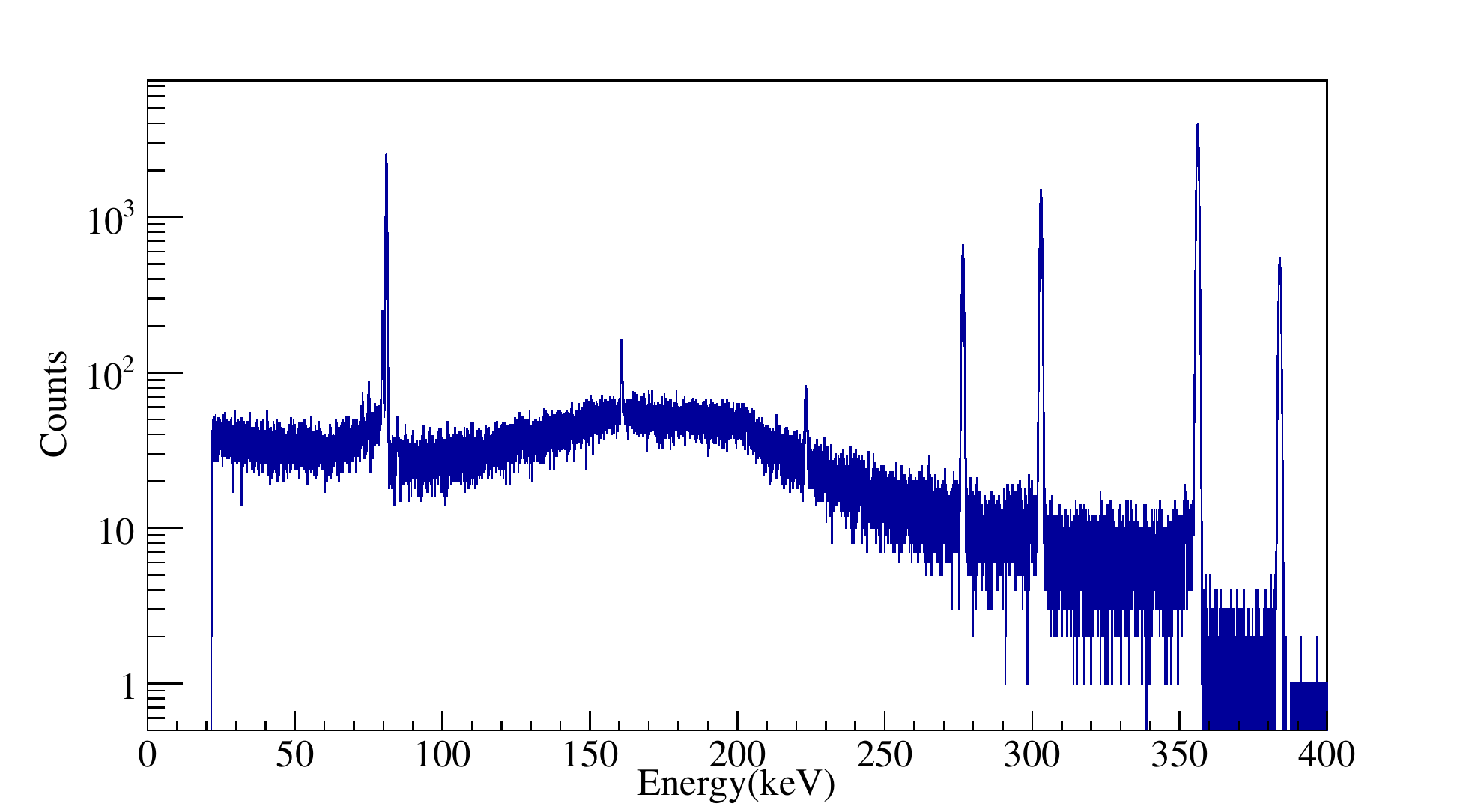}
  \figcaption{\label{fig6}An experimental spectrum of the $^{133}$Ba source by CDEX-1A PPCGe detector}
\end{center}

The fitting result of 81 keV photoelectron peak is shown in Fig.~\ref{fig7}(a). As the photoelectron peaks of 79.61 keV and 81 keV are too close to each other, two Gaussian functions and a linear function are applied to fit the two peaks and the background. The green line is the Gaussian function of 79.61 keV photoelectron peak, the blue line is the Gaussian function of 81 keV photoelectron peak, the black line is the linear function of the background and the red line is the combined results.

The fitting result of 356 keV photoelectron peak is shown in Fig.~\ref{fig7}(b). A Gaussian function and linear function are used to describe the peak region.

\begin{center}
\includegraphics[width=8.0cm,height=9.9cm]{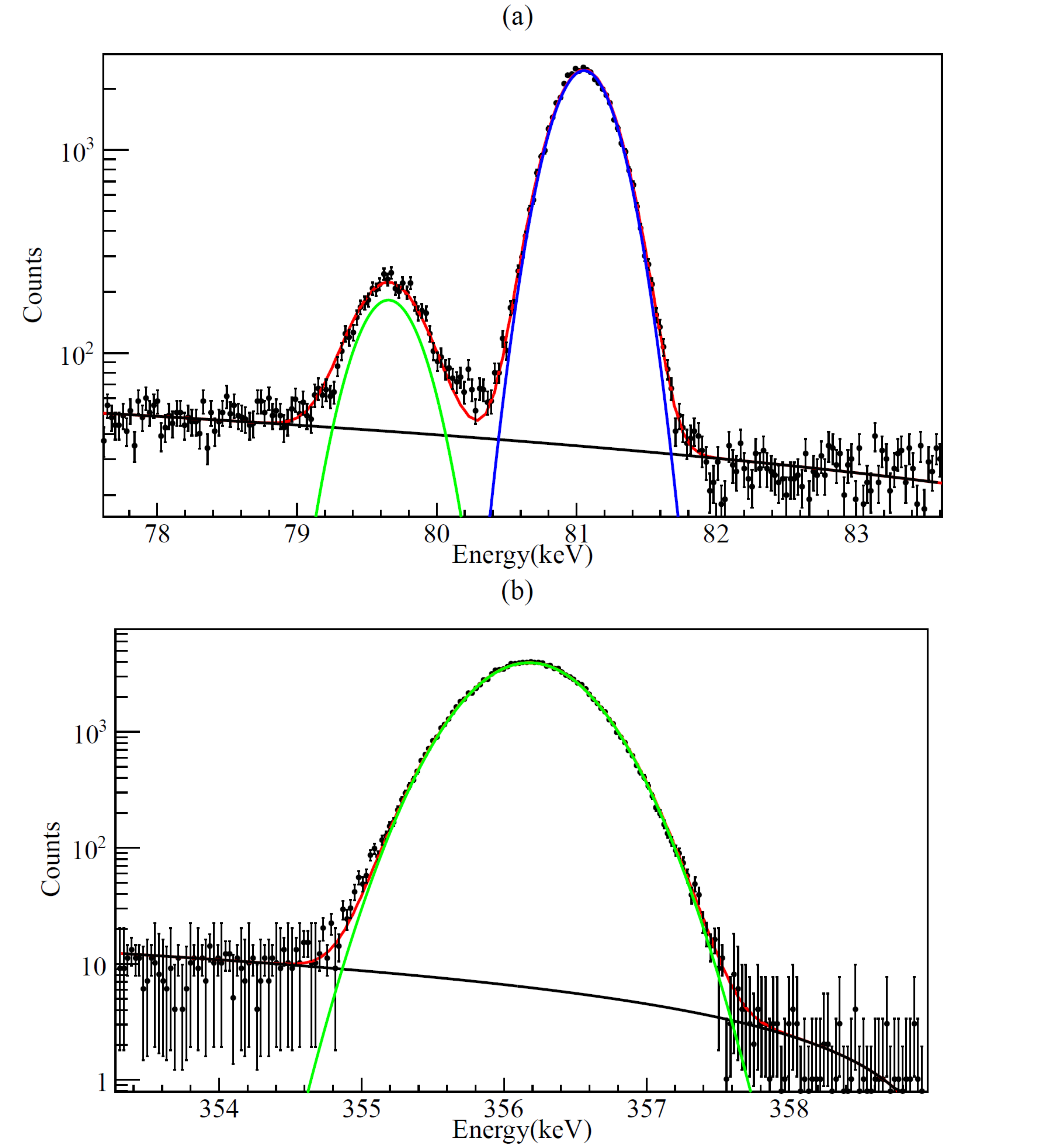}
\figcaption{\label{fig7}The fitting results of the $^{133}$Ba source. (a) The fitting result of 79.6 keV and 81 keV photoelectron peaks. (b) The fitting result of 356 keV photoelectron peak.}
\end{center}

The peak area is calculated by the formula below:

 \begin{eqnarray}
\label{eq1}
A &=& { \sqrt{2\pi}} {\sigma}{H}/{W}
\end{eqnarray}
Where A refers to the peak area of the photoelectron peak, $\sigma$ refers to the $\sigma$ of the fitting result of Gaussian function, H refers to the peak height and W refers to the width of each bin in energy.

Using the parameters derived from the fitting result of Fig.~\ref{fig7}, the ratios of different photoelectron peaks were obtained as below:

 \begin{eqnarray}
\label{eqr}
R(81keV,356keV) = \frac{A(356keV)}{A(81keV)}=\frac{{H_{356keV}}{\sigma_{356keV}}}{{H_{81keV}}{\sigma_{81keV}}}\nonumber
\end{eqnarray}

As the dead layer is at the surface of germanium detector, lower energy photons are more sensitive to it than higher energy photons. The denominator of the ratio was fixed with 81 keV peak area and the numerator was changed from 276 keV peak area to 384 keV peak area. Other experimental results were obtained by the same method and shown in Table~\ref{tab2}.

\begin{center}
\tabcaption{ \label{tab2} Different energy peak ratios with a distance of 73 mm between the source to the detector endcap}
\footnotesize
\begin{tabular*}{60mm}{c@{\extracolsep{\fill}}ccc}
\toprule Energy peak/81keV &Energy peak ratio\\
 \hline
  276keV&0.393 $\pm$ 0.005\\
  303keV& 0.953 $\pm$ 0.009\\
  356keV&2.853 $\pm$ 0.023\\
  384keV& 0.390 $\pm$ 0.005\\
\bottomrule
\end{tabular*}
\end{center}

\subsection{Simulation data analysis}

Geant4~\cite{lab13} was used to simulate the initial interaction of a $^{133}$Ba source in the PPCGe detector in CDEX-1A experiment and construct all the structures of CDEX-1A detector and shieldings into the simulation program. In the simulation, the dead layer thickness was scanned from 0 mm to 1.4 mm to get different simulation results of the $^{133}$Ba source. A simulation spectrum of $^{133}$Ba source is shown in Fig.~\ref{fig8}. As ratios are only concerned about the photoelectron peak events deposited all their energy in the bulk, surface events do not contribute at all to the photoelectron peaks no matter partial charge collected events or no charge collected events. The events in the dead layer of partial charge collection which would contribute to the low energy part of the spectrum were simplified as no charge collected events in the simulation. This is why some small photoelectron peaks can be seen in the simulation spectrum while in the experimental is not.

 \begin{center}
  \includegraphics[width=9cm,height=5cm]{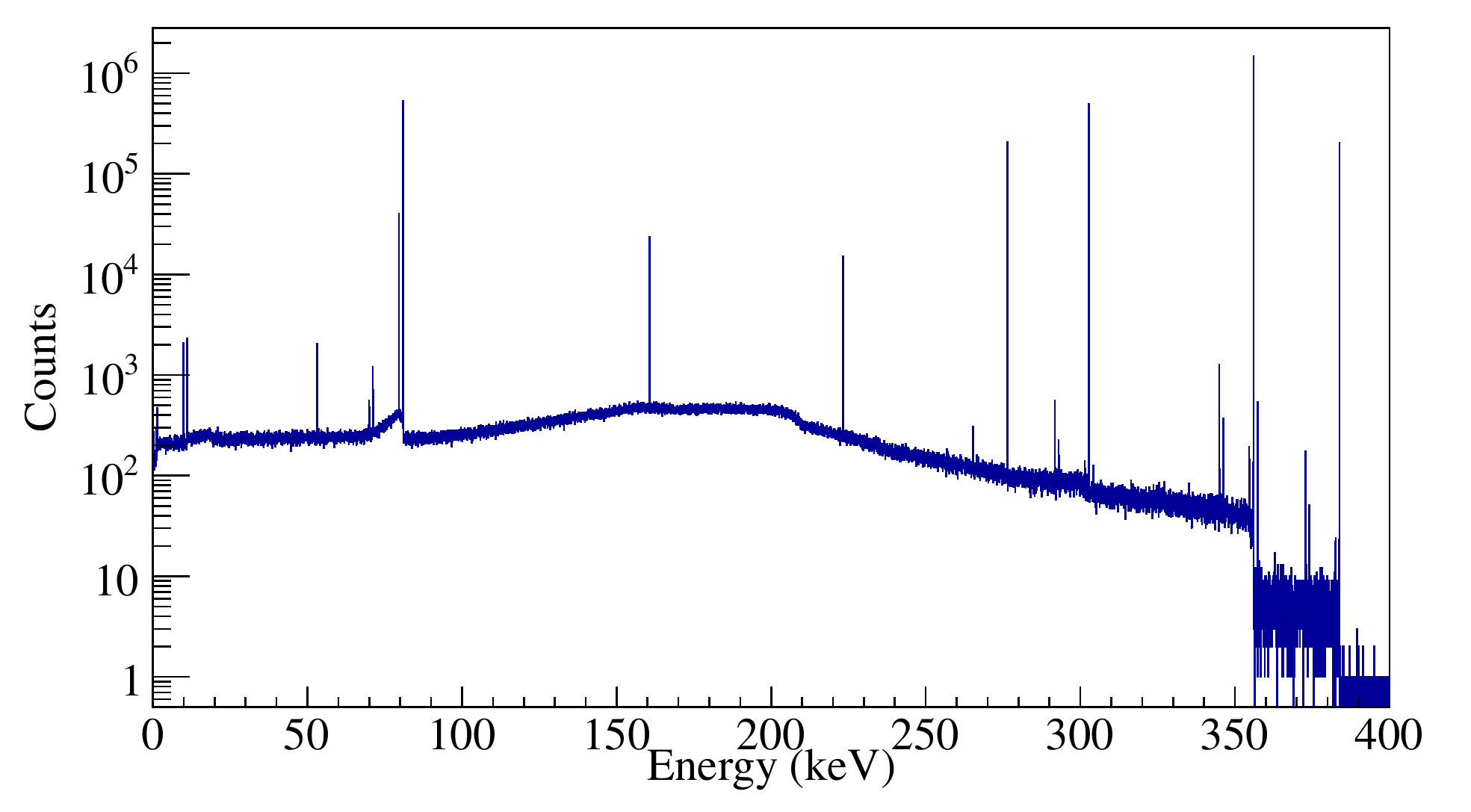}
  \figcaption{\label{fig8}The simulation spectrum of the $^{133}$Ba source. The distance between the source and endcap is 73 mm, and the dead layer thickness is assumed to be 1.0 mm.}
\end{center}

 As depicted in Fig.~\ref{fig9}, the points for different dead layer thicknesses with statistical error bars from the simulation show the ratio of the number of events in the 81 keV photoelectron peak to that in the 356 keV photoelectron peak. A quadratic fitting function provides a good description to the simulation data. The horizontal band with 1$\sigma$ error bar shows the ratio that was measured in the experimental data and the vertical band with 1$\sigma$ error bar determines the thickness of the dead layer by comparing the experimental ratio and the simulation data implementing the interpolation method. Using this method, a series of results were obtained as shown in Table~\ref{tab3}.

 \begin{center}
  \includegraphics[width=9cm,height=5cm]{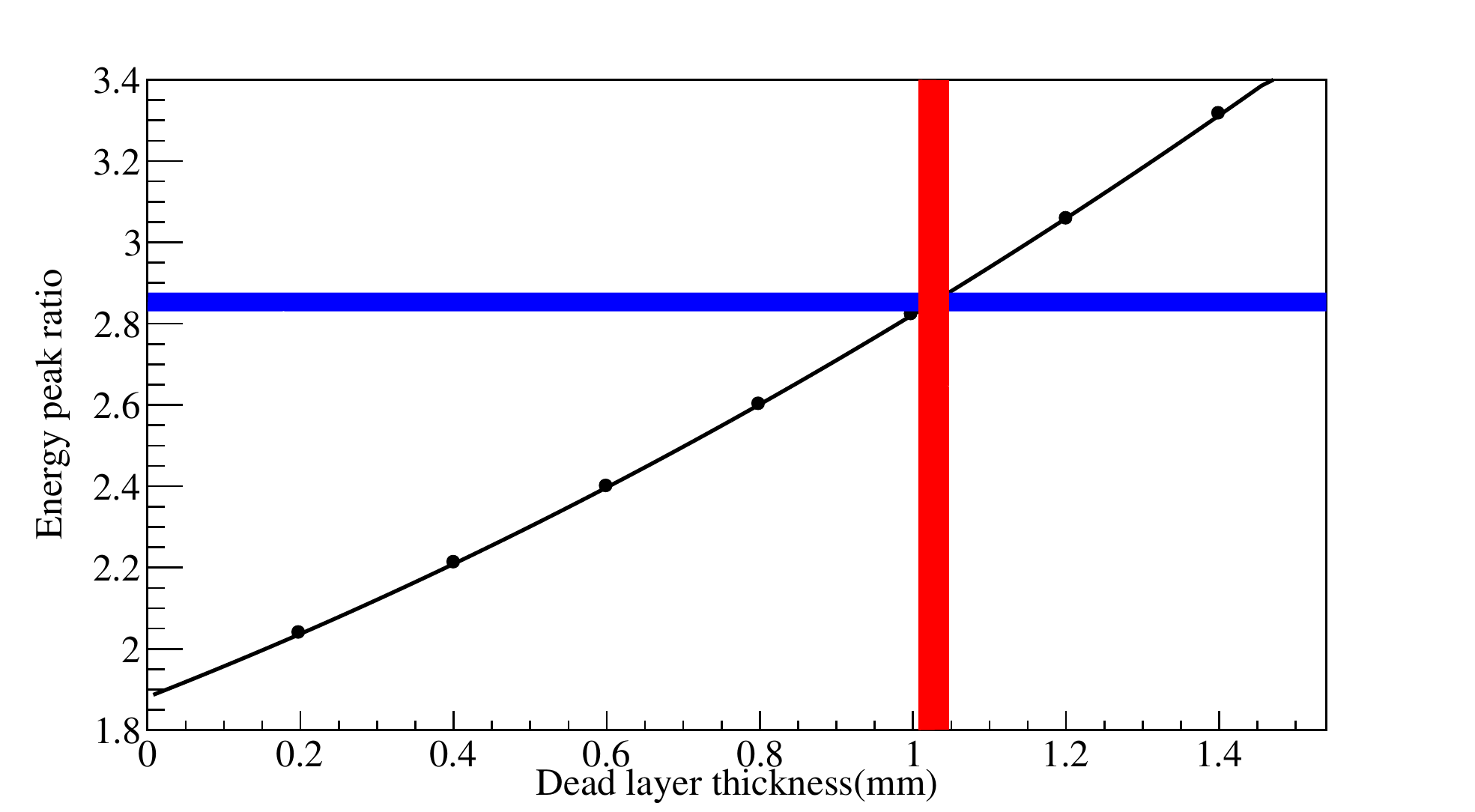}
  \figcaption{\label{fig9} Determination of the thickness of the dead layer at the source position of 73 mm. The black points are from the simulation, providing the ratio of the number of events in the 81 keV photoelectron peak to that in the 356 keV photoelectron peak. The horizontal band is from the experimental data. The vertical band determines the dead layer thickness by comparing the experimental ratio and simulation fitting line. The error bars for the simulation points are smaller than the data point size and invisible in the plot.}
\end{center}

\begin{center}
\tabcaption{ \label{tab3} The estimated dead layer thicknesses with statistical errors for different energy ratios of photoelectron peaks}
\footnotesize
\begin{tabular*}{75mm}{c@{\extracolsep{\fill}}cccc}
\toprule Enegy peaks &Dead layer thickness &Stat. error\\
 /81keV&/mm&/mm\\
  \hline
  276keV&0.994&0.033\\
  303keV& 1.035&0.025\\
  356keV&1.027&0.020\\
  384keV& 1.010&0.032\\
\bottomrule
\end{tabular*}
\end{center}

\subsection{Statistical and systematic errors}
\subsubsection{Statistical error}

In Table~\ref{tab3}, several results of the dead layer thicknesses were obtained by choosing different photoelectron peaks of $^{133}$Ba source. As the peak areas were achieved by the formula (2), the statistical errors of these results can be achieved by the following steps:

Step1: Using the formula below to get the statistical errors of different photoelectron peak ratios:

\begin{eqnarray}
\label{eq2}
\sigma_{R}=\sigma(\frac{A_{1}}{A_{2}})=\frac{H_{1}}{H_{2}}\frac{\sigma_{1}}{\sigma_{2}}\sqrt{\frac{\sigma^2_{H_{1}}}{H^2_{1}}+\frac{\sigma^2_{\sigma_{1}}}
{\sigma^2_{1}}+\frac{\sigma^2_{H_{2}}}{H^2_{2}}+\frac{\sigma^2_{\sigma_{2}}}{\sigma^2_{2}}}
\end{eqnarray}
Where $\sigma_{R}$ refers to the statistical errors of peak ratios, $A_i$ refers to the peak area of photoelectron peak i, $H_i$ refers to the peak height of photoelectron peak i, $\sigma_i$ refers to the sigma of photoelectron peak i (i=1, 2). All the errors of peak ratios were also shown in Table~\ref{tab2}.

Step2: The statistical errors of energy peak ratios were added to the experimental result as the blue band shown in Fig.~\ref{fig9}. The blue band has two intersections with the black line. A red band was derived by the two intersections in Fig.~\ref{fig9} which was the statistical error of the dead layer thickness. All the statistical errors were obtained by this method and were also shown in Table~\ref{tab3}.

Step3: From all the results in Table~\ref{tab3}, the central value and statistical error of dead layer thickness was derived with weighted average method which was 1.021 $\pm$ 0.013 mm.

\subsubsection{Systematic errors}

The systematic errors arise from:

\begin{enumerate}[(1)]
\item Events selections and rejections

The systematic error caused by events selection from preamplifier reset period was derived in three steps: Firstly, the  T$_{-}$ cut was scanned from 0.18 s to 0.22 s. Secondly, by using the same method as T$_{-}$= 0.2 s, new results from the new T$_{-}$ cut were obtained. Finally, by comparing the new results to the old result, the systematic error caused by events selection from preamplifier reset period was 0.004 mm.

The systematic error caused by accidental coincidence events rejection was derived by changing the normal signals selection region from $3\sigma$ region to $5\sigma$ region and it was 0.003 mm.

\item The accuracy of  source location and detector dimension

Other systematic errors were caused from the accuracy of source location, endcap dimension, and crystal dimensions. These systematic errors could be obtained by Monte Carlo simulations. We considered the error of endcap thickness to be 0.1 mm. Therefore, the endcap thickness was set as 1.4 mm, 1.5 mm and 1.6 mm in the simulation program, respectively. Then new dead layer thicknesses were obtained by comparing new Monte Carlo simulation results with experimental results. The difference between the dead layer thickness results we got from 1.4 mm and 1.6 mm thick endcap simulation models to the 1.5 mm thick endcap simulation model was the systematic error caused by accuracy of endcap dimension. The same method was used to get the systematic errors caused by the accuracy of source positions, and the accuracy of crystal dimensions.
\end{enumerate}

All the systematic errors were listed in Table~\ref{tab4}.

\begin{center}
\tabcaption{ \label{tab4} Systematic errors}
\footnotesize
\begin{tabular*}{85mm}{c@{\extracolsep{\fill}}cc}
\toprule
Events selection from preamplifier reset period && 0.004mm\\
Accidental coincidence events rejection && 0.003mm\\
Source location accuracy && 0.005mm \\
Endcap dimension accuracy && 0.135mm \\
Crystal dimension accuracy && 0.022mm \\
\hline
Total systematic error && 0.137mm\\
\bottomrule
\end{tabular*}
\end{center}

After all the errors including statistical error and systematic errors were considered, the dead layer thickness of the CDEX-1A PPCGe detector was derived to be 1.02 $\pm$ 0.14 mm. This result was also cross-checked by changing the $^{133}$Ba source position from 73 mm to other three positions of 42 mm, 113 mm and 159 mm. A good agreement was derived among these different source positions.

\section{Conclusion}

This paper presents the process of dead layer measurement experiment on CEDX-1A. In this study, a $^{133}$Ba source was used to measure the dead layer thickness. After all the errors were taken into account including the statistical error and systematic errors, the dead layer thickness was measured to be 1.02 $\pm$ 0.14 mm. Statistical and systematic errors were studied in detail and the endcap dimension accuracy contributed more than 90$\%$ of the total systematic error.

\acknowledgments{}

\end{multicols}

\centerline{\rule{80mm}{0.1pt}}
\vspace{2mm}

\begin{multicols}{2}

\end{multicols}

\clearpage


\begin{thebibliography}{90}

\vspace{3mm}

\bibitem{lab1}Q.Yue et al.,Phys. Rev. D,2014,{\bf90}:091701(R).
\bibitem{lab2}P. N. Luke et al.,IEEE Trans. Nucl. Sci.,1989,{\bf36}:926---930.
\bibitem{lab3}W. Zhao et al.,Phys. Rev. D,2013,{\bf88}:052004.
\bibitem{lab4}LV Zi-feng et al.,Chinese Physics C,2012,{\bf36(9)}:855---860.
\bibitem{lab5}C. E. Aalseth et al.,Phys. Rev. Lett.,2011,{\bf106}:131301.
\bibitem{lab6}E. Aguayo et al.,Nucl. Instr. and Meth. A,2013,{\bf701}:176---185.
\bibitem{lab7}J. L. Campbell et al.,Nucl. Instrum. Methods,1974,{\bf117}:519---532.
\bibitem{lab8}P. A. Burns et al.,Nucl. Instrum. Meth. A,1990,{\bf286}:480---489.
\bibitem{lab9}A. Clouvas et al.,Health Physics,1998,{\bf74}:216---230.
\bibitem{lab10}N. Q. Huy et al.,Nucl. Instrum. Meth. A,2007,{\bf573}:384---388.
\bibitem{lab11}Nuclear Data Sheets,2011,{\bf112}:935---936.
\bibitem{lab12}M. Agostini et al., Journal of Instrumentation,2011,{\bf6}:P04005.
\bibitem{lab13}http://geant4.web.cern.ch/geant4/.




\end{thebibliography}
\end{document}